\documentclass[12pt]{article}
\usepackage{scicite}
\usepackage{times}
\usepackage{graphicx}
\usepackage{epsfig}
\usepackage{amsmath}
\usepackage[usenames,dvipsnames]{color}
\usepackage[export]{adjustbox}
\usepackage{bbold}
\usepackage{braket,amsfonts}
\usepackage{mathrsfs}
\usepackage{amsmath}
\usepackage{marvosym}
\usepackage{amssymb}
\usepackage{algorithm}
\usepackage[noend]{algpseudocode}
\usepackage{amsfonts}
\usepackage{xr}
\usepackage{caption}
\usepackage{hyperref}

\hypersetup{
    colorlinks=true,
    linkcolor=blue,
    filecolor=magenta,      
    urlcolor=cyan,
    pdftitle={Overleaf Example},
    pdfpagemode=FullScreen,
    }

\newenvironment{sciabstract}{
\begin{quote} \bf}
{\end{quote}}
\newcounter{lastnote}

\newcommand{\real}{{\mathbb{R}}}

\newcommand{\x}{\operatorname{x}}
\newcommand{\ctrl}{\operatorname{u}}
\newcommand{\ctrlvec}{\underline{\bf u}}
\newcommand{\xvec}{\underline{\bf x}}
\newcommand{\lvec}{\underline{\bf \lambda}}
\newcommand{\zvec}{\underline{\bf z}}
\newcommand{\vvec}{\underline{\bf v}}
\newcommand{\xivec}{\underline{\boldsymbol{\xi}}}
\newcommand{\lambdavec}{\underline{\boldsymbol{\lambda}}}
\newcommand{\nuvec}{\underline{\boldsymbol{\nu}}}
\newcommand{\Bvec}{\underline{\bf B}}

\newcommand{\sym}{\operatorname{Sym}}

\title{Optimal control of interacting active particles \\ on complex landscapes}

\author{S. Sinha$^{1, \dagger}$, V. Krishnan$^{1, \dagger}$,    \& L. Mahadevan$^{1,2\ast}$
\\
\footnotesize{$^1 $ School of Engineering and Applied Sciences, Harvard University, Cambridge, MA 02138, USA}\\
\footnotesize{$^2$  Departments of Physics, and Organismic and Evolutionary Biology, Harvard University, Cambridge, MA 02138, USA}\\
\footnotesize{$^\ast$To whom correspondence should be addressed; E-mail:  lmahadev@g.harvard.edu}\\
\footnotesize{$^\dagger$Equal Contribution}
}

\date{}
\begin{document}

\baselineskip16pt
\maketitle

\begin{sciabstract}
 Active many-body systems composed of many interacting degrees of freedom often operate out of equilibrium, giving rise to non-trivial emergent behaviors which can be functional in both evolved and engineered contexts. This naturally suggests the question of \emph{control}  to optimize function. Using navigation as a paradigm of function, we deploy the language of stochastic optimal control theory to formulate the inverse problem of shepherding a system of interacting active particles across a complex landscape. We implement a solution to this high-dimensional problem using an Adjoint-based Path Integral Control (APIC) algorithm that combines the power of recently introduced continuous-time back-propagation methods and automatic differentiation with the classical Feynman-Kac path integral formulation in statistical mechanics. Numerical experiments for controlling individual and interacting particles in complex landscapes show different classes of successful navigation strategies as a function of landscape complexity, as well as the intrinsic noise and drive of the active particles. However, in all cases, we see the emergence of paths that correspond to traversal along the edges of ridges and ravines, which we can understand using a variational analysis. We also show that the work associated with optimal strategies is inversely proportional to the length of the time horizon of optimal control, a result that follows from scaling considerations. All together, our approach serves as a foundational framework to control active non-equilibrium systems optimally to achieve functionality, embodied as a path on a high-dimensional manifold.
\end{sciabstract}

\newpage

The emergence of complex patterns in space-time from simple interactions between particles is a major theme in statistical and continuum physics of non-equilibrium active matter \cite{MRS-AW-RGW-GG-HR:20}. This is most clearly evident in biology where manifestations of morphological and functional complexity abound, and are present across scales. The emergence of collective functional physiology and behavior, which allows living systems to be viable, requires control and regulation \cite{NW:19}. This necessitates a shift in perspective from the forward problem of determining the evolution of patterns given the rules, 
the conventional paradigm in physics, towards the need for the solution of the inverse problem of determining the control to evoke functionality \cite{AR-NW-JB:43,GFRE-JK:19,HR:00}. 
 
Here we frame and solve this inverse problem in the context of a specific function, guidance and control of a system of active interacting particles on a complex landscape, with the goal of moving them from an initial to a target configuration, subject to control costs. The abstract paradigm of navigation on manifolds has many concrete realizations: the control of interacting spins in either a classical or quantum setting \cite{PU:21, MCE-JAS-MPB:23, JW-EKUG:07}, the folding of proteins to create allostery \cite{PGW-JNO-DT:95,WZ-BRB-DT:06, JWR-NP-IB-CPG-AJL-SRN:17 }, speed and trajectory of evolution \cite{SI-ED-JC-JP-NK-OG-BKS-SD-EI-JGS:21}, the control of soft robotic systems \cite{DAHH-MJB-EK-ABC-PCC-IM-EWH:23}, the design of meta-materials \cite{AAZ-MJM-LV-JBH:23,MS-AM:23,DRR-NP-JMW-HMJ-AJL-SRN-JJDP:18,NP-DH-AJL-SRN:19}, the control of  collectives (of insects, robots, and other interacting agents) \cite{DR-MTT:15,SGP-SM-FG-JK-VNM-LM:22} , and learning in deep neural networks \cite{YB-JK-JP-SSS-JSD-SD:20}. All these problems can be mathematically characterized as systems with finite degrees of freedom (DOFs) governed by many-body stochastic dynamics, with internal driving and/or the ability to control and steer them via an external field~\cite{DGO-CJ-CAP-AP-ET:22, LP-ET-RG:21, BL-HL:19}. The functional efficacy of the controlled system can be measured in terms of the cost of control that has two contributions, a running cost and a terminal cost that measures the accuracy of achieving the target. The problem of optimal control is then to find the guidance strategy that minimizes the total cost subject to the constraints imposed by the dynamical system. This is an inherently non-equilibrium, nonlinear, stochastic problem which also suffers from the curse of dimensionality associated with the 
exponential dependence of the volume of the search space of solutions on the number of DOFs~\cite{RB:57}.
But as we will see, recent advances in the practical ability to solve forward and backward problems using automatic differentiation \cite{AEB-YCH:75, AEB-WFD:62, DER-GEH-RJW:86,AGB-BAP-AAR-JMS:18} and thus propagate variations in the solutions to parameter changes allows us to combine ideas from stochastic optimal control theory \cite{RFS:86,HK:05,ET-ET:12,WHF-SKM:82} and machine learning \cite{PK:22, jax2018github} to make progress. 

For concreteness, we consider a system of $N$ active particles whose controlled dynamics is described by the overdamped Langevin equation,
\begin{align} \label{eq:N-body_system}
     \dot{\xvec}(t) = -\frac{1}{\eta}\nabla V(\xvec(t)) + \ctrlvec(t) + \sqrt{2D} \xivec(t), \qquad \xvec(0) = \xvec_0
\end{align}
where $\xvec(t)= ( {\bf x}_1(t), {\bf x}_2(t),...{\bf x}_N(t) )$ is the configuration (i.e, positions ${\bf x}_i(t) \in \real^d$) of the $N$ particle system at time $t$ interacting via a dynamic potential $V(\xvec(t))$, subject to an additive control $\ctrlvec(t)$ that needs to be determined via some extraneous condition, and   $\xivec(t)$ is a delta-correlated, stationary Gaussian process at time $t$, with statistics given by $\langle {\xi}_i^{\alpha} \rangle$=0, and $\langle { \xi}_i^{\alpha}(t) {\xi}_j^{\beta}(t')\rangle=\delta(t-t')\delta^{\alpha \beta}\delta_{ij}$, where $\alpha=1,2,..d$ and $i,j \in \{ 1, \ldots, N \}$,
and the gradient operator $\nabla = \frac{\partial}{\partial \xvec}$. The parameters~$\eta$ and~$D$ in~\eqref{eq:N-body_system} are the damping and  diffusion constants respectively.  While all our numerical results are associated with assuming that the ambient dimension~$d=2$, our framework is valid for arbitrary dimension and our equations are generalizable to their appropriately invariant forms on curved manifolds. In Eqn. \eqref{eq:N-body_system}, we assume overdamped dynamics and hence neglect inertia, and further postulate that since we are working with non-equilibrium systems, $\eta$ and $D$ need not satisfy the fluctuation-dissipation relation \cite{JPH-IRM:13}; this naturally allows us to also consider the athermal limit, $D=0$ corresponding to the deterministic limit \cite{SFE-RBSO:89, JWR-NP-IB-CPG-AJL-SRN:17}. For simplicity, we choose all the particles to be identical, and further that the friction factor $\eta=1$ unless specified otherwise. The control task involves steering the system across the dynamic landscape generated by $V$ for a time interval, $[0,T]$, with the requirement of minimizing the work done by the controller and with the goal of minimizing the target error $\Psi(\xvec(T))$, on the terminal state, $\xvec(T)$. The control task can be formulated as the following stochastic optimization problem
\begin{align} \label{eq:control_task}
    \min_{\ctrlvec_{[0,T]}} ~ \mathbb{E}_{\mathbb{Q}_{\ctrlvec_{[0,T]}}} \left[ \left. \frac{\gamma}{2} \int_0^T \left\| \ctrlvec (t') \right\|^2 dt' + \Psi(\xvec(T)) ~\right|~  \xvec(0) = \xvec_0 \right] 
\end{align}
subject to the dynamics given by Eqn. (\ref{eq:N-body_system}). In Eqn. (\ref{eq:control_task}), the expectation 
\sloppy $\mathbb{E}_{\mathbb{Q}_{\ctrlvec_{[0,T]}}}[G(\xvec)]=\int_{\xvec(0)=\xvec_0}~\mathcal{D}\xvec~ G(\xvec)~\mathbb{Q}_{\ctrlvec_{[0,T]}}(\xvec)$, is over the distribution of all paths $\mathbb{Q}_{\ctrlvec_{[0,T]}}$ generated by \eqref{eq:N-body_system} with the control $\ctrlvec_{[0,T]}$
over the time interval $[0,T]$ and $\mathcal{D}\xvec$ is an infinitesimal volume element in the space of paths, and the condition $\xvec(0) = \xvec_0$, and the parameter~$\gamma$ captures the relative weight assigned to the total work done, $W=\int_0^T \left\| \ctrlvec \right\|^2(t') dt'$ compared to the terminal penalty, $\Psi(\xvec(T))$.

Using a standard approach first proposed by Bellman~\cite{RB:57}, 
we introduce a value function $F(t,\zvec)$ that denotes the optimal cost-to-go at time~$t$ from state $\xvec(t) = \zvec$ and is defined as
\begin{align}
    F(t, \zvec) = \min_{\ctrlvec_{[t,T]}} ~\mathbb{E}_{\mathbb{Q}_{\ctrlvec_{[t,T]}}} \left[ \left. \frac{\gamma}{2} \int_t^T \left\| \ctrlvec (t') \right\|^2 dt' + \Psi(\xvec(T)) ~\right|~  \xvec(t) = \zvec \right]
    \label{value_fn}
\end{align}
and reduce the global formulation above to a local condition known as the the \emph{Hamilton-Jacobi-Bellman (HJB)} equation \cite{RB:57}, a nonlinear partial differential equation (PDE) given by (see SI section S2 for a brief derivation)
\begin{align} \label{eq:stochastic_HJB}
    \frac{\partial F}{\partial t} + D \Delta F  -  \nabla F \cdot \nabla V - \frac{1}{2\gamma} \left| \nabla F \right|^2 = 0
\end{align}
along with the temporal boundary condition $F(T,\zvec) = \Psi(\zvec)$, i.e. it must be solved in backward time from $t=T$. The HJB equation~\eqref{eq:stochastic_HJB} suffers from the curse of dimensionality~\cite{RB:57} owing to the nonlinearity and the state space being of dimension~$N >> 1$. To circumvent this, we first reformulate it using the Cole-Hopf transformation $F(t,\zvec) = - \frac{1}{\beta} \log \varphi (t, \zvec)$, for some $\beta > 0$, and then use our freedom in the choice of the parameter $\beta$ by imposing the relation $\beta = \frac{1}{2\gamma D}$ (analogous to the Stokes-Einstein relation), to reduce the HJB equation to a linear advection-diffusion equation~\cite{HK:05,HJK:05} (also in backward time):
\begin{align} \label{eq:HJB_cole_hopf}
    \frac{\partial \varphi}{\partial t} + D \Delta \varphi -  \nabla \varphi \cdot \nabla V  = 0,
\end{align}
with the boundary condition $\varphi(T, \zvec) = e^{-\beta \Psi(\zvec)}$.
The solution to the transformed equation~\eqref{eq:HJB_cole_hopf} 
can be expressed as a path integral using the Feynman-Kac formula~\cite{PDM:04}
\begin{align} \label{eq:FK_formula_varphi}
     \varphi(t, \zvec) =  \mathbb{E}_{\mathbb{P}_{[t,T]}} \left[ \left. e^{-\beta \Psi(\xvec(T))}  \; \right| \;
        \dot{\xvec}(t') = - \nabla V(\xvec(t')) + \sqrt{2D} \xivec(t'), ~ \xvec(t) = \zvec \right].
\end{align}
An important consequence of this approach is that the expectation above is taken with respect to the distribution 
$\mathbb{P}_{[t,T]}$ of paths generated by the \emph{uncontrolled dynamics} $\dot \xvec(t') = -\nabla V(\xvec(t') + \sqrt{2D}\xivec(t')  $ over the time interval $t' \in [t,T]$ starting at $\zvec$, i.e., satisfying the condition $\xvec(t) = \zvec$. Correspondingly, the value function, $F(t,\zvec)$,  can be expressed as
\begin{align} \label{eq:value_func_PI}
    F(t,\zvec) = - \frac{1}{\beta} \log \left( \mathbb{E}_{\mathbb{P}_{[t,T]}} \left[ \left. e^{-\beta \Psi(\xvec(T))} \; \right| \;
        \dot{\xvec}(t') = - \nabla V(\xvec(t')) + \sqrt{2D} \xivec(t'), ~ \xvec(t) = \zvec  \right] \right)
\end{align}
and the optimal control $\ctrlvec^*(t) = - \frac{1}{\gamma} \nabla F(t, \xvec(t))$. It is not coincidental that the value function~\eqref{eq:value_func_PI} 
takes the form of a free energy, wherein the (path integral) partition function corresponds to a Boltzmann distribution over uncontrolled paths, with the weight factor $e^{-\beta \Psi(\xvec(T))}$ obtained from the cost of an uncontrolled path $\Psi(\xvec(T))$. All together, we note that the value function for the optimal control can be
computed entirely from the uncontrolled dynamics, a property that follows from two assumptions~\cite{HK:05,HJK:05} (i) the control $\ctrlvec$ enters linearly in the dynamics
and (ii) the cost is quadratic in $\ctrlvec$. While this might seem limiting, there are many problems where these assumptions are entirely reasonable; furthermore, in the context of modern approaches to the control of stochastic systems using model-predictive-control (MPC), these assumptions can be iteratively used to achieve good approximations to more complex forms of the controller and associated costs.

To solve the combination of the forward and backward (adjoint) problems iteratively and derive the optimal control strategy, we now describe an easily implementable Adjoint-Path Integral Control Algorithm.  The path integral representation in \eqref{eq:value_func_PI} yields the optimal control at time $t$ as $\ctrlvec^*(t) = - \frac{1}{\gamma} \nabla F(t, \xvec(t))$ and involves propagating the state $\xvec(t)$  via the uncontrolled dynamics over the time interval $[t,T]$  to a new state $\xvec(T)$ at which point the gradient is evaluated, carried out via adjoint method, formulated next. From~\eqref{eq:FK_formula_varphi} and~\eqref{eq:value_func_PI},
we first note that the gradients of~$F(t,\zvec)$ and~$\varphi(t,\zvec)$ are related by 
\begin{align} \label{eq:F_phi_gradient_relationship}
 \nabla F(t,\zvec) = - \frac{\nabla \varphi(t,\zvec)}{\beta \varphi(t,\zvec)}.    
\end{align}
To compute the gradient of $\varphi(t,\zvec)$ with respect to $\zvec$ in \eqref{eq:FK_formula_varphi}, we note that the uncontrolled dynamics $\dot{\xvec}(t') = - \nabla V(\xvec(t')) + \sqrt{2D} \xivec(t')$ must be propagated starting from $\zvec$ at time $t$ (i.e., $\xvec(t) = \zvec$) over the time interval~$[t,T]$.
Using the calculus of variations,  we treat the uncontrolled dynamics as a constraint to construct the Lagrangian for the evaluation of the expectation in~\eqref{eq:FK_formula_varphi}
\begin{align*}
    \mathcal{L}(t, \xvec, \nuvec) =  \mathbb{E}_{\xivec} \left[ e^{-\beta \Psi(\x(T))} - \int_t^T \nuvec^\top \left( \dot{\xvec}(t')  + \nabla V(\xvec(t')) - \sqrt{2D} \xivec(t') \right) dt' \right],
\end{align*}
where $\nuvec$ is the Lagrange multiplier for the constraint of uncontrolled dynamics and the expectation above is with respect to 
the stochastic process~$\xivec$. From setting the first variation with respect to $\xvec$ in the interval $(0,T)$ and the terminal $\xvec(T)$ above to zero, we obtain the
stationary conditions which fix the evolution of the Lagrange multiplier $\nuvec$ as follows
\begin{align}
\begin{aligned}
    &\dot{\nuvec}(t') = \nabla^2 V(\xvec(t')) \nuvec(t'), \\
    \text{subject~to} \quad &\nuvec(T) = - \beta e^{-\beta \Psi(\xvec(T))} \nabla \Psi(\xvec(T)).
\end{aligned}
\end{align}
 Recalling the relation between the gradients of~$F$ and~$\varphi$ given by \eqref{eq:F_phi_gradient_relationship} (see SI for details), and making a change of variables $\lambdavec(t') = - \frac{ \nuvec(t') }{\beta \mathbb{E}_{\xivec} \left[ e^{-\beta \Psi(\xvec(T))} \right]  },$
we  obtain the adjoint equations for the evolution of the scaled Lagrange multiplier as follows
\begin{align} \label{eq:adjoint_PI}
\begin{aligned}
    \dot{\lambdavec}(t') = \nabla^2 V(\xvec(t')) \lambdavec(t'), \qquad 
    \lambdavec(T) = \frac{e^{-\beta \Psi(\xvec(T))}}{\mathbb{E}_{\xivec} \left[ e^{-\beta \Psi(\xvec(T))} \right]} \nabla \Psi(\xvec(T)), 
\end{aligned}
\end{align}
where $\nabla^2 V$ is the Hessian (matrix) of potential $V$ (with $\nabla^2 = \frac{\partial^2}{\partial \xvec_i \partial \xvec_j}$) and the optimal control at time~$t$ is then given by
\begin{align} \label{eq:optimal_ctrl_adjoint}
    \ctrlvec^*(t) = - \frac{1}{\gamma} \left. \nabla F(t,\zvec) \right|_{\zvec = \xvec(t)} = - \frac{1}{\gamma} \mathbb{E}_{\xivec} \left[ \lambdavec(t) \right]
\end{align}
For a practical implementation of the algorithm, as outlined in the algorithm above, we see from~\eqref{eq:adjoint_PI} that  to evaluate the optimal control $\ctrlvec^*$ at time~$t$, we first sample several instances of the uncontrolled forward dynamics from the current state~$\xvec(t)$ up to time~$T$ for different realizations of the stochastic process~$\xivec$, evaluate the gradient~$\nabla \Psi$ at~$\xvec(T)$ for each sampled path and back-propagate~$\lambdavec$ from the terminal condition in~\eqref{eq:adjoint_PI} for each sample to evaluate the average~$\lambdavec$ across the samples at time~$t$. We note that the Boltzmann weight factor in the value function~\eqref{eq:value_func_PI} essentially reduces to a relative weighting of the terminal gradient of the sampled uncontrolled paths,
as specified in~\eqref{eq:adjoint_PI}. Furthermore, note that the Hessian of the potential~$V$ along the uncontrolled path governs the adjoint back-propagation in~\eqref{eq:adjoint_PI}. 

Before proceeding to discuss our numerical results, we note that the athermal limit of the optimal control can be obtained by setting~$D=0$ in the HJB equation~\eqref{eq:stochastic_HJB} and computing its gradient with respect 
to~$\zvec$, to obtain
\begin{align}
    \frac{\partial}{\partial t} \nabla F  
    -  \nabla^2 F\left(  \nabla V + \frac{1}{\gamma} \nabla F \right) 
    - \nabla^2 V \nabla F = 0,
\end{align}
where $\nabla^2 F$ is the Hessian of $F$. The optimal trajectory $\xvec^*(t)$ 
satisfies $\dot{\xvec}^*(t) = - \nabla V(\xvec^*(t)) + \ctrlvec^*(t)
= - \nabla V(\xvec^*(t)) - \frac{1}{\gamma} \nabla F(t, \xvec^*(t))$,
and we see that 
\begin{align}
    \frac{D}{D t} \nabla F (t, \xvec^*(t)) - \nabla^2 V(\xvec^*(t)) \nabla F(t, \xvec^*(t)) = 0.
\end{align}
Letting $\lambdavec^*(t) = \nabla F(t, \xvec^*(t))$ in the equation above,
we get $\dot{\lambdavec}^*(t) = \nabla^2 V(\xvec^*(t)) \lambdavec^*(t)$
and the boundary condition $\lambdavec^*(T) = \nabla F(T, \xvec^*(T)) = \nabla \Psi(\xvec^*(T))$ to obtain the following equations of motion for the system $(\xvec^*(t), \lambdavec^*(t))$~\cite{VGB-RVG-YEFM-LSP:62}
\begin{align} \label{eq:pontryagin}
\begin{aligned}
    \dot{\xvec}^*(t) &= -\nabla V(\xvec^*(t)) + \ctrlvec^*(t), \qquad \xvec^*(0) = \xvec_0 \\
    \dot{\lambdavec}^*(t) &= \nabla^2 V(\xvec^*(t)) \lambdavec^*(t), \qquad \quad ~\lambdavec^*(T) = \nabla \Psi(\xvec^*(T)) \\
    \ctrlvec^*(t) &= - \frac{1}{\gamma} \lambdavec^*(t)
\end{aligned}
\end{align}
This is equivalent to constructing a Hamiltonian $\mathcal{H}(\zvec, \vvec)
 = \frac{\gamma}{2} \left\| \vvec \right\|^2 + \lambdavec^\top \left( -\nabla V(\zvec) + \vvec \right)$ that allows us to rewrite the optimal solution as 
$\dot{\lambdavec}^*(t) = - \nabla \mathcal{H}(\xvec^*(t), \ctrlvec^*(t), \lambdavec^*(t))$
and $\ctrlvec^*(t) = \arg \min_{\vvec} \mathcal{H}(\xvec^*(t), \vvec, \lambdavec^*(t))$.
Thus we see that the optimal control minimizes the Hamiltonian at every point along the optimal trajectory, provides an alternative to the local HJB equation, and is the well-known \emph{Pontryagin Minimum Principle}~\cite{VGB-RVG-YEFM-LSP:62}. This allows us to interpret \eqref{eq:pontryagin} as the Lagrangian formulation
for the corresponding Eulerian HJB formulation (in the terminology of continuum mechanics).

In our implementation, the (first and higher order) derivatives of the interaction potential are computed by automatic differentiation within the framework supplied by JAX~\cite{jax2018github}. High-dimensional ODE and SDE integrations are performed in Diffrax~\cite{PK:22}, a JAX-based library for numerical integration of differential equations. 
For the implementation of the sampling-based Feynman-Kac path integral adjoint algorithm, we utilize the automatic vectorization functionality provided by ${\rm vmap}$ in JAX in combination with ODE/SDE integration in Diffrax.

Before describing the results of our numerical experiments with the APIC algorithm, we use a variational argument to suggest the emergence of a universal strategy. For simplicity, we start with the athermal limit of the stochastic optimal control problem (i.e. Eqn. \ref{eq:control_task} ) which then reduces to



\begin{align} 
    \min_{\ctrlvec} ~\frac{\gamma}{2} \int_0^T \left\| \ctrlvec \right\|^2(t) dt + \Psi(\xvec(T)),
    \qquad \text{s.t.} \quad \dot{\xvec}(t) = -\nabla V(\xvec(t)) + \ctrlvec(t); \quad \xvec(0)=\xvec_0
    \label{athermal_eq}
\end{align}
Substituting $\ctrlvec(t) = \dot{\xvec}(t) + \nabla V(\xvec(t))$ into the functional (\ref{athermal_eq}) yields the following unconstrained minimization problem for $\xvec$:
\begin{align*}
   \mathcal{L}= &\min_{\xvec} ~\frac{\gamma}{2} \int_0^T \left\| \dot{\xvec}(t) + \nabla V(\xvec(t)) \right\|^2 dt + \Psi(\xvec(T)) \\
    &= \min_{\xvec} ~ \frac{\gamma}{2} \int_0^T \left\| \dot{\xvec}(t) \right\|^2 dt 
    + \gamma \int_0^T \nabla V(\xvec (t))^{T} \cdot \dot{\xvec}(t) dt
    + \frac{\gamma}{2} \int_0^T \left\| \nabla V(\xvec(t)) \right\|^2 dt 
    + \Psi(\xvec(T)).
\end{align*}
 Setting  $\delta \mathcal{L}/\delta\xvec=0 $ yields following  equation of motion for the optimal trajectory of the state $\xvec(t)$, given by
\begin{align}
    \ddot{\xvec}(t) = \nabla^2 V(\xvec(t)) \nabla V(\xvec(t)) = \frac{1}{2} \nabla \left( \left| \nabla V(\xvec(t)) \right|^2 \right)
    \label{emer_eqn}
\end{align}
with $\xvec(0) = \xvec_0$ and the terminal boundary condition $\dot{\xvec}(T) = - \nabla V (\xvec(T)) - \frac{1}{\gamma} \nabla \Psi(\xvec(T))$.

The physical interpretation of the \textit{emergent} Eqn.~\eqref{emer_eqn} is that in flat regions of the landscape, the particle follows the simple geodesic straight-line path ($\ddot{\xvec} = 0$) whereas in non-flat regions it accelerates in the direction of increasing magnitude of the gradient of~$V$. Furthermore, the behavior is symmetric under the transformation $V \rightarrow -V$ of the potential, which implies that hills and valleys of the landscape have the same effect. 

%

 To verify the predictions of the theoretical analysis, we performed four numerical experiments  with (a) a single stochastic particle in a frozen landscape of hills and valleys,  (b) Interacting stochastic particles in a frozen landscape, and  (c) Interacting athermal particles. To create the mixed landscape, the static potential was chosen as 
\begin{align}
   V_{\rm s}(\mathbf{r}) = \sum_{i=1}^n \eta_i \Theta \left( \left( \left\| \mathbf{r} - \mathbf{x}_i^s \right\| - d_0 \right)^2 \right),
\end{align}
where $\Theta$ is the Heaviside step function, the number of modes in the frozen landscape $n = 80$, $\eta_i$ was sampled uniformly from the interval $(-0.01, 0.01)$, 
$\mathbf{x}_i^s$ were sampled uniformly from a square of side $0.8$ units and $d_0 = 0.04$. In the numerical simulations, the time step for numerical integration by the Euler-Maruyama method was chosen to be $dt = 0.01$ and the number of sampled paths in the implementation of the APIC algorithm was $20$. The parameters $\gamma = 1$, $D= 5 \times 10^{-4}$ and $T=3$. 

Figure \ref{fig:hills_valleys}A shows the trajectories of individual particles, shown as tracks in different colors, moving in a mixed landscape of hills (red) and valleys (yellow). Consistent with the above analysis (see Eqn. \ref{emer_eqn}), 
particles follow straight-line paths ($\ddot{\xvec} = 0$) in flat regions, while in non-flat regions they accelerate in the direction of increasing magnitude of the gradient of~$V$, owing to which they tend to skirt the ridges or valleys of the landscape. Furthermore, the hills and valleys of the landscape appear to have the same effect on the behavior of the particles. Refer to Movie1 (see details in the SI section S7) for the animation of the numerical experiment \cite{SS-VK-LM:23_movies}.

Separately, we also considered a set of stochastic non-interacting particles on a stationary landscape consisting only of valleys (Figure S1 in the SI) to closely investigate their behavior in the vicinity of varying potential (as if zoomed in on a portion of Figure~\ref{fig:hills_valleys}A). The landscape consisted of a Gaussian mixture potential, 
$V_{\rm s}({\bf r})=\frac{\nu}{\sqrt{2\pi \sigma^2}} \sum_{i=1}^{i=N}e^{-\frac{||{\bf x}_i^s-{\bf r}||^2}{2\sigma^2}}$, where ${\bf x}_i^s$ is the location of the $i^{th}$ static particle generated from a uniform distribution, with $N=20$, $\sigma^2=0.1$, and $\nu=-0.01$. Five non interacting particles were initialized in a square box of size $0.05$ centered at $(-0.4, -0.4)$ and the control task was to steer the particles to the origin, $(0,0)$, while minimizing the total work done. Figure S1 in the SI shows that optimal particle trajectories follow the ridges of the landscape in agreement with the theoretical prediction Eqn. \eqref{emer_eqn}. Refer to Movie6 (see details in the SI section S7) for the animation of the numerical experiment \cite{SS-VK-LM:23_movies}.

Next, we consider the control of interacting particles connected by springs in a static external landscape implemented using the same Gaussian mixture potential as above.  The inter-particle potential is assumed to be of the form $V(\xvec)=\frac{1}{2}\sum_{i=1}^{N}\sum_{j\in NN(i)}k(||{\bf x}_i-{\bf x}_j||-l)^2$, where $N=5$, $k=0.1$, and $l=0.2$. the nearest-neighbor of the ith particle, $NN(i)$, is based on a distance cut-off of 1.5 units. In Fig. \ref{fig:hills_valleys}B and C, we show the plots of optimal trajectory of the center-of-mass (COM) of the particles for different parameter values, and in lower right inset of Fig. \ref{fig:hills_valleys}B (C) we show the value of the optimal control for $D=0.0001, K=0.0001$ ($D=0.01, K=0.1$). Again, we see that the COM traverses the ridge between the two deep valleys of the static landscape, albeit in a stochastic manner. The shape of the five particle network is remarkably different for the two cases at the end of the trajectory. In the case of softly connected particles in the low-noise limit ($D=0.0001, K=0.0001$) the particles remain coherent, whereas they spread when the diffusivity and stiffness are larger ($D=0.01, K=0.1$). Interestingly, there is a significant difference in the final profile of the particles; for low diffusion/stiffness $D=0.0001, K=0.0001$, the profiles are similar to the non-interacting case except when moving through the ridge   (see upper and lower right corners of Figure \ref{fig:hills_valleys}C), while in the strongly interacting noisy limit ($D=0.01, K=0.1$),  the control varies erratically   (see upper and lower right corners of Figure \ref{fig:hills_valleys}C). Refer to Movie2 (see details in the SI section S7) for the animation of the numerical experiment \cite{SS-VK-LM:23_movies}.


A natural question that these numerical experiments raise is that of a general strategy for the guided navigation of interacting particles from one location to another. Inspired by recent work on the control of active droplets in the athermal limit \cite{SS-VR-LM:22}, that suggest a gather-move-spread solution, we next considered a set ($N=30$) of interacting athermal particles ($D=0$), using the same potential as before, with the initial positions of the particles sampled from a uniform distribution with support $[-3, 3] \times [-3, 3]$ (see Figure \ref{fig:gather_spread}A for the initial configuration). The task was to reach a prescribed boundary corresponding to a circle of radius 4 units at time~$T$. Fig. \ref{fig:gather_spread}B, \ref{fig:gather_spread}C, and \ref{fig:gather_spread}D show the trajectories of individual particles in numerical experiments with different time-horizons $T=0.1, 0.5,$ and $1$ respectively, corresponding to short or long times compared to collective intrinsic time scale in the uncontrolled problem given by $\sim 0.1$ time units (extracted as fit to potential energy relaxation in Fig. \ref{fig:gather_spread}E). For $T=0.1$, the mechanical network just spreads out and individual particles move along straight lines; for $T=0.5$, the mechanical network initially comes together and then spreads outwards, while for $T=1$, the mechanical network first shrinks collectively and then spreads outwards. This gather-spread strategy can be rationalized by considering the potential energy, $V_{\rm int}(\xvec(t))$ shown in Fig. \ref{fig:gather_spread}E. For $T=0.1$, the potential increases monotonically from the initial non-equilibrium state, while for $T=0.5$, the potential initially decreases (`gather' phase) from the non-equilibrium state and subsequently increases (`spread' phase), and finally for $T=1$, the potential energy first relaxes to zero from the initial non-equilibrium state (`gather' phase), and then increases (`spread' phase) to fulfill the task. Refer to Movie3 (see details in the SI section S7) for the animation of the numerical experiment \cite{SS-VK-LM:23_movies}.

We note that the relaxation of the potential energy to zero is consistent with our earlier argument that the system as a whole prefers regions of flat landscapes. Fig. \ref{fig:gather_spread}F, \ref{fig:gather_spread}G, and \ref{fig:gather_spread}H show the controls for $T=0.1, 0.5$, and $T=1$, respectively, and shows that the controls for the case $T=0.1$ act from the start, in contrast to the case of $T = 0.5$ and $T = 1$, where the initial magnitude of the control is negligible. Furthermore, the value of the control is highest for $T=0.1$, consistent with the idea that a larger time horizon might be a better way to control many body systems to allow one to exploit the intrinsic (uncontrolled) dynamics for as long as possible. 


To quantify the dependence of the total work done $W=\int_{0}^{T}||\ctrlvec||^2 dt$ as a function of the changing time-horizon, $T$, we note that since $\ctrlvec$ has dimensions of velocity ($\frac{[L]}{[T]}$), $[W]\sim \frac{[L]^2[T]}{[T]^2}\sim \frac{[L]^2}{[T]}$, i.e $W\sim \frac{1}{T}$.  To test this, we considered the deterministic optimal control of a single particle to the origin as it moves in a complex Gaussian mixture landscape given by $V_{\rm s}({\bf r})=\frac{\nu}{\sqrt{2\pi \sigma^2}} \sum_{i=1}^{i=N}e^{-\frac{||{\bf x}_i^s-{\bf r}||^2}{2\sigma^2}},$ where ${\bf r}$ is a spatial location in two-dimensions, ${\bf x}_i^s$ is the location of the $i^{th}$ static particle, generated from a uniform distribution, with $N=20$, $\sigma^2=0.1$, and $\nu=-0.01$. The control cost is assumed to be given by $\min_{\ctrlvec} ~ \mathcal{C}:=\int_0^T \big[\frac{\gamma}{2}  \left\| \ctrlvec \right\|^2(t)+ \beta V_{\rm s}(\xvec(t))\big] dt + \Psi(\xvec(T)),$ where $\beta=2$.  
The main panel of Figure \ref{fig:det_single_land}A shows the particles trajectory for T=[0.01, 0.05, 0.1, 0.12]; when $T=0.01$, the trajectory of the particle is straight whereas the curvature in the trajectory increases as $T$ increases, consistent with our previous arguments. In the top inset of \ref{fig:det_single_land}A), we see that as the time $T$ increases, so does the control $||\ctrlvec||$, and the work done, $W$ is inversely proportional to the time horizon, $T$, as shown in the lower inset of \ref{fig:det_single_land}A), consistent with our scaling argument. Refer to Movie4 (see details in the SI section S7) for the animation of the numerical experiment \cite{SS-VK-LM:23_movies}.

Finally, we briefly discuss the tension between extrinsic noise strength $D$ and intrinsic energy scale encoded in $V$, using the same interaction potential as before, in the context of a task to drive the COM of the mechanical network to the origin on a flat landscape. Fig.~\ref{fig:det_single_land}(B,C) show the  trajectories of COM overlaid with snapshots of the shape of mechanical network for $D=0.01, K=0.0001$ and $D=0.01, K=1.0$, respectively. Consistent with intuition, the mechanical network spreads a lot for $D=0.01, K=0.0001$ (low stiffness) in comparison to $D=0.01, K=1.0$ (high stiffness). Refer to Movie5  (see details in the SI section S7) for the animation of the numerical experiment \cite{SS-VK-LM:23_movies}.

 
We have shown how a combination of concepts from optimal control theory and statistical mechanics along with efficient computational methods provides a framework for the control of functional many-body systems, instantiated in terms of the navigation of a system of particles. Augmenting the dynamics of the system using co-state variables for optimal control and using the Feynman-Kac path integral in a computationally tractable framework that uses automatic differentiation (JAX) for the efficient solution of the forward-backward adjoint problem leads to our A(djoint) based P(ath) I(integral) Control algorithm. Deploying this in a range of situations for interacting particles moving on complex landscapes shows how particles prefer to move along ridges, while the optimal strategies take advantage of the intrinsic dynamics of the particles when the time horizon is large, leading to a gather-move(-spread) strategy, with the sparing use of control only when necessary. We further see that the work done by the controller is consistent with a simple scaling law, and that there is a tradeoff between control of center of mass and variance of the position as a function of interaction strength and noise. Although our study has focused on the continuous-time control of discrete particulate systems, the same approach can be generalized to the case of continuous media in the context of optimal transport~\cite{JDB-YB:00,SS-VR-LM:22} as well as discrete-time control of both particulate/agent-based models~\cite{ET:06,HJK-VG-MO:12} and continuous media~\cite{RJ-DK-FO:98}, objects of study for the future. Finally, our framework can be used to study the evolutionary dynamics of populations by allowing the landscape itself to change (slowly) in response to the movement of the particles on it, and introducing a selection principle for the population.   
    
{\bf Acknowledgments.} We thank NSF grants  BioMatter DMR 1922321, MRSEC DMR 2011754, EFRI 1830901,  ONRG N629092012026, the Simons Foundation and the Henri Seydoux Fund, the Dana Farber Cancer Institute (Prof. C. Z. Zhang), and the Wellcome Trust for partial financial support.

\newpage

\bibliography{references,snake}
\bibliographystyle{Science}

\newpage

\begin{figure}
\includegraphics[width=\columnwidth]{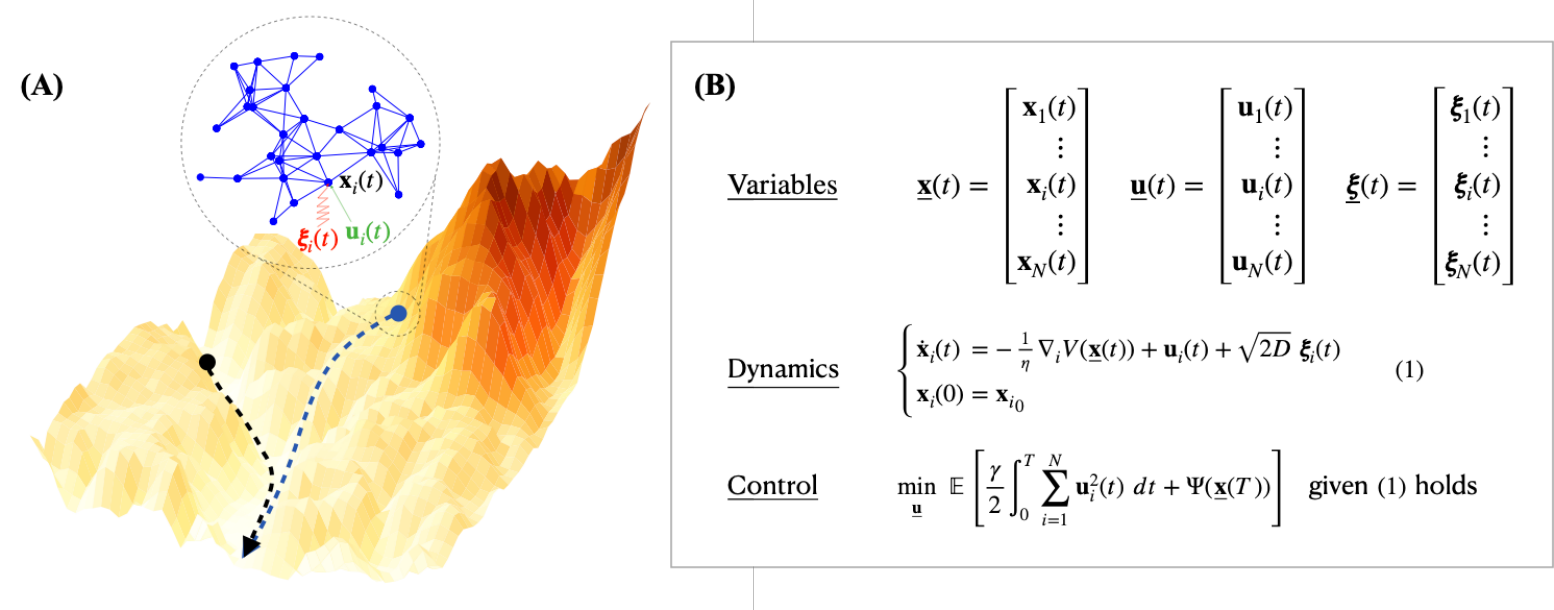}
\caption{\footnotesize{ \bf Schematic for navigation of active systems and the stochastic optimal control formulation.}
{\bf (A)} Navigation of an interacting active particle system (blue network of nodes and edges) on a landscape. In this schematic, the rugged terrain depicts the complex energy landscape, which the active particle system is navigating. The blue network is a representation of the interacting particle system, where the blue nodes are particles (position depicted by ${\bf x}_i(t)$) and the edges denote the interaction between the particles. The control ${\bf u}_i(t)$, drives the active particle system on the rugged landscape, which gives rise to distinct non-equilibrium phases. The noise of the environment is modeled by a delta-correlated white noise, $\boldsymbol{\xi}_i(t)$.
{\bf (B)} Mathematical description of the stochastic optimal control for interacting systems at finite effective temperature. The system undergoes overdamped Langevin dynamics under the influence of potential, $V({\bf x}_1, {\bf x}_2, ...., {\bf x}_N)$, delta-correlated white noise $(\boldsymbol{\xi}_1, \boldsymbol{\xi}_2, \ldots, \boldsymbol{\xi}_N)$, and control $({\bf u}_1, {\bf u}_2, \ldots, {\bf u}_N)$. The control task is then to minimize the expectation ($\mathbb{E}$) of a running cost $\int_{0}^{T}\sum_{i=1}^{N} \left\| {\bf u}_i (t) \right\|^2 dt$, and a terminal cost $\Psi({\bf x}_1(T), {\bf x}_2(T)....{\bf x}_N(T))$, where the expectation is with respect to controlled trajectories. In the current study, we also formulate the optimal navigation for athermal, $D=0$, many-body interacting systems. In this limit, the stochastic optimal control reduces to the deterministic~case.} 
\label{fig:landscape_ctrl}
\end{figure}

\begin{figure}[!ht]
\centering
\includegraphics[width=0.55\linewidth]{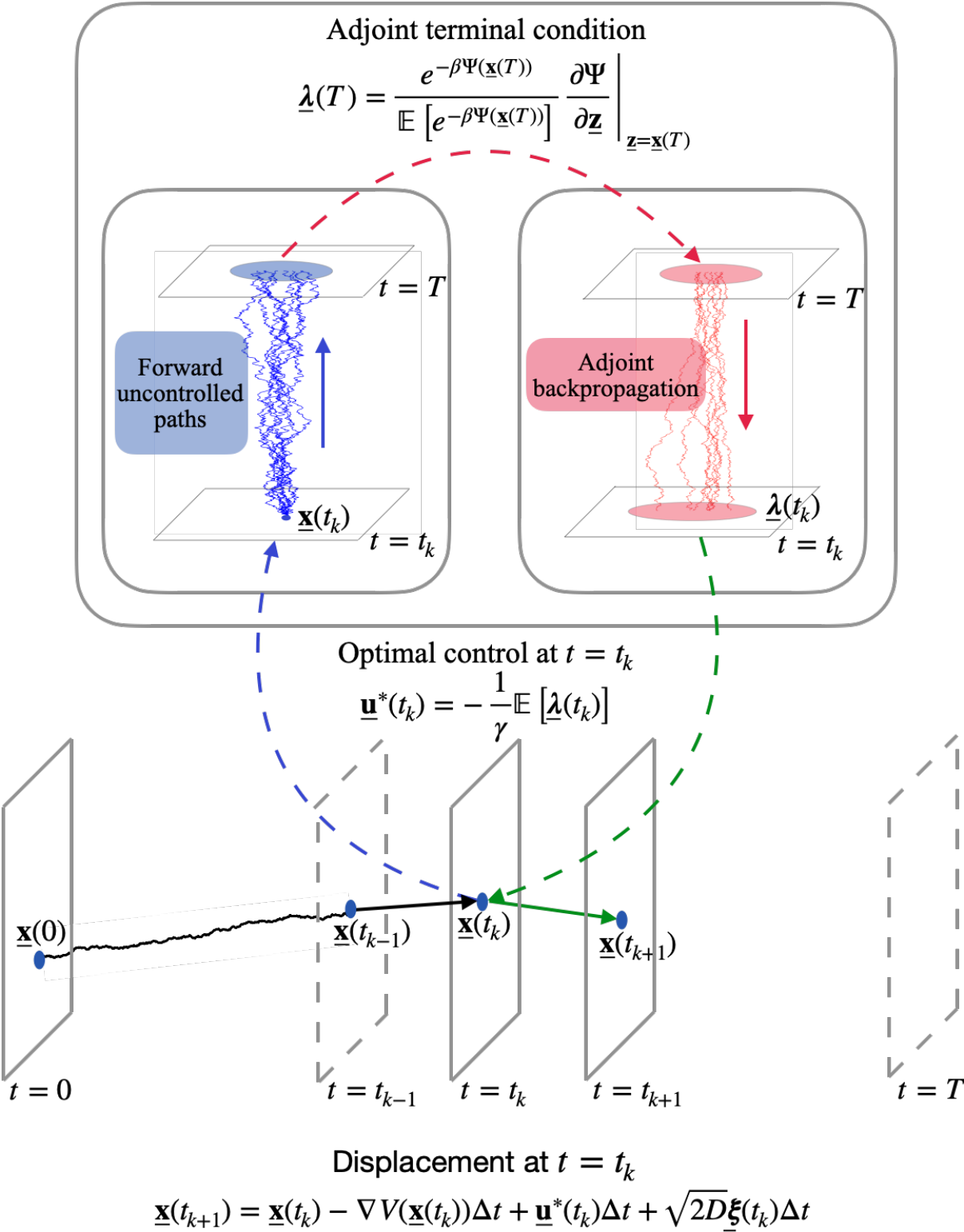} \hspace{0.03\linewidth}
\begin{minipage}{0.40\linewidth}
\vspace{-5.0in}
\begin{algorithm}[H] 
\caption*{\footnotesize{\bf Adjoint Path Integral Control Algorithm}}
\footnotesize
\textbf{Input:} State $\xvec(t)$, Number of paths~$n$ \\
\textbf{For every time $t$:}
\begin{algorithmic}[1]
	\State Obtain $n$ independent Brownian noise sequences for the interval $t_0$ to $T$
	\State Integrate SDE~\eqref{eq:N-body_system} with generated Brownian noise sequences to obtain $n$ uncontrolled paths with initial condition $\xvec(t)$
	\State Integrate adjoint ODE~\eqref{eq:adjoint_PI} with terminal condition for $\lambdavec(T)$ obtained from the terminal state $\xvec(T)$ of each uncontrolled path to obtain $n$ samples of~$\lambdavec(t)$
        \State Obtain optimal control $\ctrlvec^*(t)$ from~\eqref{eq:optimal_ctrl_adjoint} by sample averaging
\end{algorithmic}
	\label{alg:}
\end{algorithm}
\end{minipage}
\caption{\footnotesize{ {\bf  Adjoint method for stochastic optimal control based on the Feynman-Kac path integral formalism.}
{\bf (Left panel)} Schematic illustrating the method for finite-time-horizon stochastic optimal control. A trajectory is shown with time progressing horizontally. The inset shows the computational step at time $t_0$, $0\leq t_0 \leq T$, to obtain the optimal control input $\ctrlvec^*(t_0)$. The adjoint method uses the Feynman-Kac path integral to propagate several uncontrolled paths 
(solutions to $\dot{\xvec}(t) = -\nabla V(\xvec(t)) + \sqrt{2D} \xivec(t)$), $t \in [t_0,T]$, with initial condition $\xvec(t_0)$ with independently sampled Brownian noise sequences. The resulting terminal states $\xvec(T)$, set the terminal conditions on the (adjoint) co-state $\lambdavec(T)$,
as the Boltzmann-weighted gradient of the terminal cost $\Psi$ evaluated at $\xvec(T)$, which are then backpropagated through the adjoint ODE $\dot{\lambdavec}(t) =  \nabla^2 V (\xvec(t)) \lambdavec(t)$ backwards in the domain $t \in[T,t_0]$, resulting in a back-propagated adjoint path for each forward uncontrolled path. The optimal control $\ctrlvec^*(t_0)$ is then obtained from the sample average of the back-propagated adjoint co-states $\lambdavec(t_0)$.
Computationally, we use automatic differentiation within the framework supplied by JAX~\cite{jax2018github} to compute derivatives and Diffrax~\cite{PK:22}, a JAX-based library for numerical integration of differential equations. 
For the implementation of the sampling-based Feynman-Kac path integral adjoint algorithm, we utilize the automatic vectorization functionality provided by ${\rm vmap}$ in JAX in combination with ODE/SDE integration in Diffrax.
{\bf (Right panel)} Table containing the Adjoint-based Path Integral Control algorithm. } }
\label{fig:adjoint}
\end{figure}

\begin{figure}
\centering
\includegraphics[width=0.65\linewidth]{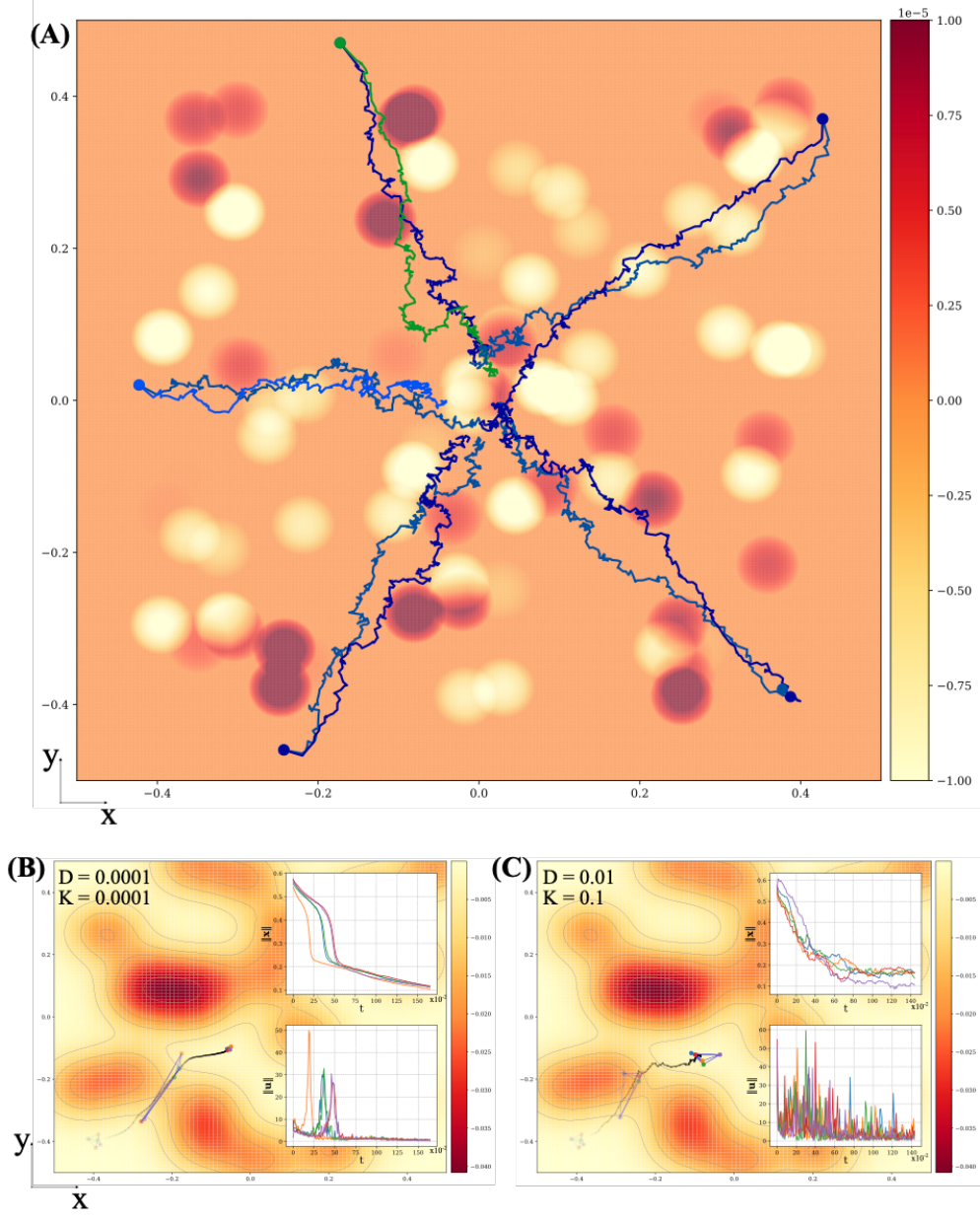}
\caption{\footnotesize{{\bf Stochastic optimal control of particles in a frozen landscape.} (A) The optimal task is to drive the particles to the origin. Trajectories of particles, shown as tracks in different colors (the colors of the trajectories are only for illustration), moving in a mixed landscape of hills (red) and valleys (yellow). The color scale on the right shows the value of the frozen landscape. In accordance with the theory (see Eqn. \eqref{emer_eqn}), the particles move along ridges. 
The parameters $\gamma = 1$, $D= 5 \times 10^{-4}$ and $T=3$. The static potential was chosen as $V_{\rm s}(\mathbf{r}) = \sum_{i=1}^n \eta_i \Theta \left( \left( \left\| \mathbf{r} - \mathbf{x}_i^s \right\| - d_0 \right)^2 \right)$,
where $\Theta$ is the Heaviside step function, the number of modes in the frozen landscape $n = 80$,
$\eta_i$ was sampled uniformly from the interval $(-0.01, 0.01)$, 
$\mathbf{x}_i^s$ was sampled uniformly from a square of side $0.8$ units and $d_0 = 0.04$.
In the numerical simulations, the time step for numerical integration by the Euler-Maruyama method was chosen to be $dt = 0.01$ and the number of sampled paths in the implementation of the Adjoint-based PI control was $20$. {\bf (B)} Plot shows the trajectory of COM of five interacting particles moving in a Gaussian mixture landscape for $D=0.0001$ and $K=0.0001$. The top-right inset shows the distance of the particles from the origin, and the bottom-right shows the magnitude of the control for all the particles. {\bf (C)} Same as in {\bf (B)} but for $D=0.01$ and $K=0.1$. In both cases,
the system of interacting particles prefer to avoid regions of low potential, in accordance with theory given by Eqn.~\eqref{emer_eqn}. }}
\label{fig:hills_valleys}
\end{figure}

\begin{figure}[!ht]
\centering
\includegraphics[width=1.0\linewidth] {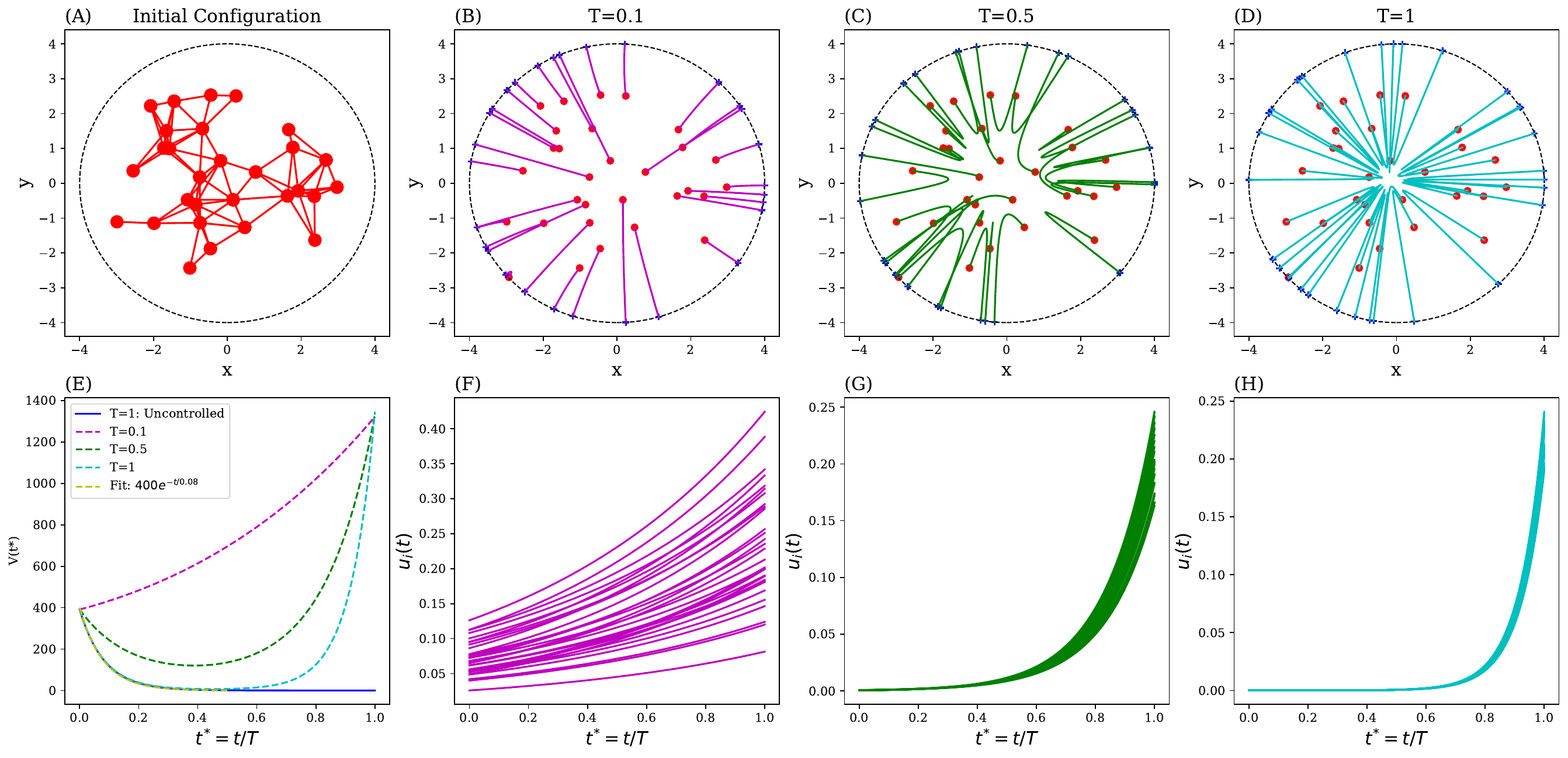}
\caption{\footnotesize{{\bf Time-horizon induced emergence of gather-spread strategy in an active network of non-linear springs.} In this control task, all the particles of the network are supposed to migrate to a circle of radius 4 units.} {\bf (A)} Initial configuration of a non-linear spring network, where the node is a particle's position, and the edge is the non-linear spring interaction. The potential energy, V,  is given by, $V({\bf x}_1,...,{\bf x}_N)=\frac{1}{2}\sum_{i=1}^{N}\sum_{j\in NN(i)}k(||{\bf x_i}-{\bf x_j}||-l)^2$, where N=30, k=0.1, l=0.2. The initial $x$ and $y$ coordinates of the particles are sampled from a uniform distribution with support from [-3, 3]. The nearest-neighbor of the $i^{th}$ particle, $NN(i)$, is based on a distance cut-off of 1.5 units, and remains the same throughout the experiment. {\bf (B)} The trajectories of the particles for time-horizon, T=0.1. In this case, the trajectories are predominantly straight lines, which does not exhibit gather-spread strategy. {\bf (C)} The trajectories of the particles for time-horizon, T=0.5. In this case, the network initially shrinks (i.e gather phase) and then migrates to the circle. This exhibits relatively weak gather-move-spread strategy. {\bf (D)} Trajectories corresponding to T=1, exhibiting strong gather-move-spread strategy. In this case, the network initially shrinks to a small radius and then spreads to the outer circle. {\bf (E)} Plot of potential energy, V, as a function of scaled time, $t^{*}=\frac{t}{T}$. As before, time has been scaled to plot the curves corresponding to different $T$, on the same graph. The uncontrolled system (green line) undergoes relaxation with a characteristic time-scale of approximately 0.08 units (shown as an exponential fit in cyan). Potential energy for $T=0.1$, shows a monotonic increase throughout the experiment corresponding to just the `spread' phase. Potential energy for both $T=0.5$ (blue) and $T=1$ (red) undergo an initial decrease, corresponding to the `gather' phase, and then a monotonic increase corresponding to `spread' phase. The gather spread in T=1 case is more pronounced than T=0.5 case, as the potential energy approximately decreases to zero for the former. {\bf (F, G, H)} The individual controls, $u_i$, for $T=0.1, 0.5$ and $T=1$. Controls for the case T=0.1, start acting from the start which is in stark contrast to the case of $T=0.5$ and $T=1$, where the initial magnitude of the control is negligible. Also, the dispersion in the values of control is highest for T=0.1, implying that there exists high degree of variability in potential energy barriers to reach the outer circle. The maximum value of the control is also highest for $T=0.1$, implying that higher time-horizon might be a better way to control many body systems as the internal dynamics can be exploited. }
\label{fig:gather_spread}
\end{figure}

\begin{figure}
\centering
\includegraphics[width=0.58\linewidth]{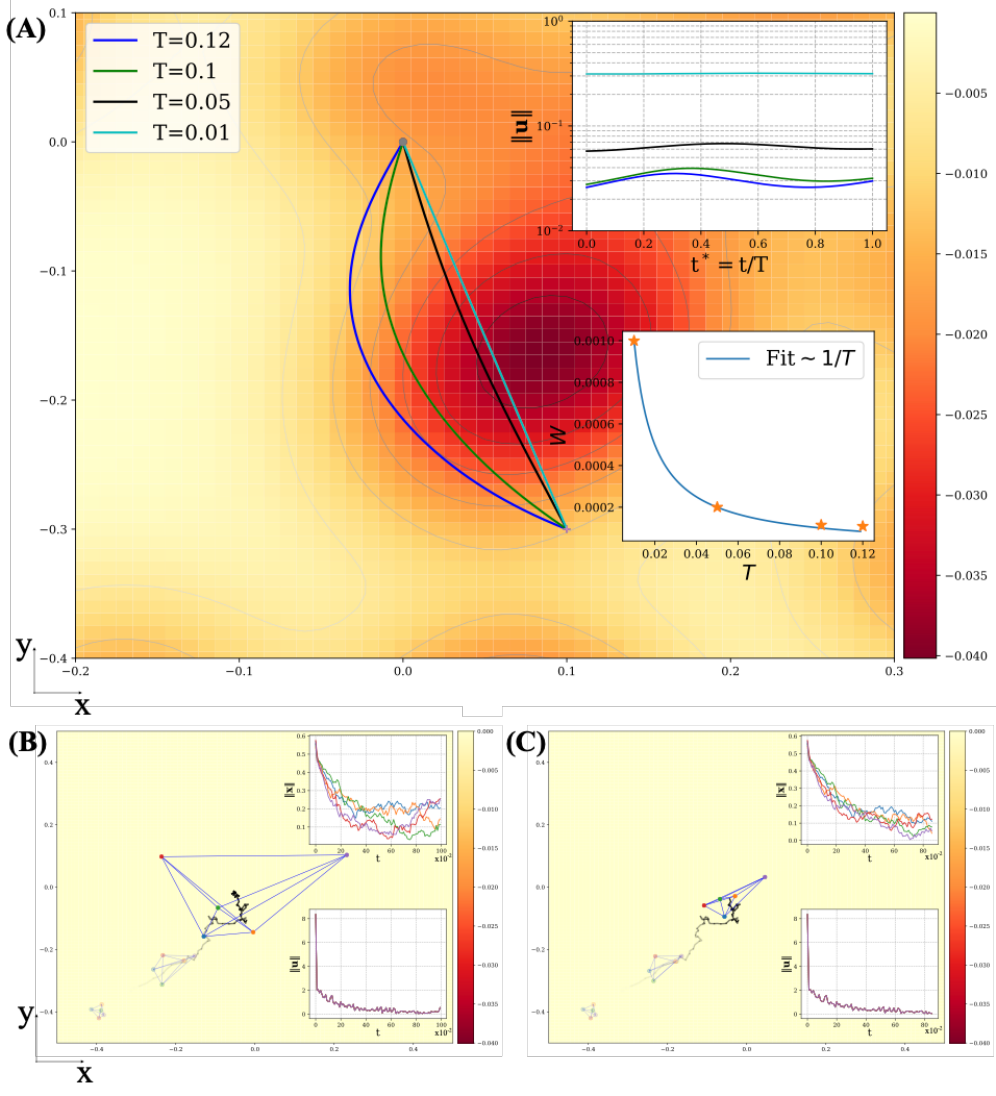}
\caption{\footnotesize{{\bf Path of single particles moving on a frozen landscape.} (A) $V({\bf r})=\nu \sum_{i=1}^{i=N}\frac{1}{\sqrt{2\pi D}}e^{-\frac{||{\bf x}_i^s-{\bf r}||^2}{2D}}$, where {\bf r} is a spatial location in two-dimensions, ${\bf x}_i^s$ is the location of the $i^{th}$ static particle, generated from a uniform distribution, which generates the Gaussian potential, $N=20$, $2D=0.1$, and $\nu=0.01$. Regions of high potential are blue in color, whereas the yellow are regions of minimal potential. In this control task, the particles underwent athermal dynamics under the influence of the frozen landscape. The control task was to move from the initial location (0.1, -0.3), to final point (0, 0). The task was chosen as the initial and final locations are separated from a potential barrier. From left to right, the curves correspond to particle trajectories corresponding to different time-horizons: $T=0.12, T=0.1, T=0.05,$ and $T=0.01$. The plot shows as the the time horizon is increased, the particle undergoes curved trajectories to escape regions of high potential. Inset on the upper-right shows control magnitude, $|{\bf u}|$, as a function of scaled time, $t^{*}=\frac{t}{T}$, for the numerical experiment in the main figure on a log-linear plot. The time has been scaled so that experiments corresponding to different time-horizons can be plotted on the same graph. From bottom to top, the curves correspond to controls for T=0.12, 0.1, 0.05, 0.01. The control for T=0.12 is approximately two orders of magnitude smaller than T=0.01. Inset on the lower-right shows the magnitude to total work, $W=\int_{0}^{T}||\ctrlvec||^2 dt$, as a function of time-horizon, $T$. The orange stars correspond to the numerical experiment, whereas the blue line corresponds to the fit, $W\sim \frac{1}{T}$, estimated from theory. {\bf (B, C)} Stochastic optimal control of interacting active particles in the absence of a static external landscape. The particles were initialized in a square box of size $0.05$ centered at $(-0.4, -0.4)$ and the control task was to steer the particles to the origin $(0,0)$ while minimizing the total work done.  Stochastic optimal control of five interacting particles without a landscape. {\bf (B)} Plot shows the trajectory of the center-of-mass (COM) (in black), of the five interacting particles for $K=0.0001$ and $D=0.01$. Overlaid on the trajectory are snapshots of the five particles (in red, with the network edges representing the interactions) at different time instances. The top-right inset shows the distance of the particles from the origin, and the bottom-right shows the magnitude of the control for all the particles (hard to distinguish). {\bf (C)} Same as in {\bf (B)} but with $K=1.0$ and $D=0.01$.}}
\label{fig:det_single_land}
\end{figure}

\clearpage
\section*{\LARGE Supplementary Information (SI)}

\appendix

\setcounter{figure}{0}  
\renewcommand{\thesection}{S\arabic{section}}
\renewcommand{\theequation}{S.\arabic{equation}}
\renewcommand{\thefigure}{S\arabic{figure}}

\section{Mathematical notation}
\begin{itemize}
    \item $\xvec(t)$: The configuration of the $N$ particle system (i.e position), at time $t$. 
    \item $\dot{\xvec}(t)$: The velocity~$\frac{d\xvec(t)}{dt}$.
    \item $\nabla$: Gradient operator = $\frac{\partial}{\partial \xvec}$.
    \item $\nabla^2$: Hessian operator = $\frac{\partial^2}{\partial \xvec_i \partial \xvec_j}$.
    
    \item $V$: Scalar potential function dictating the interactions. In the present study, $V=V_{\rm s}+V_{\rm int}$, where $V_{\rm s}$ is the external static landscape and $V_{\rm int}$ is the interactions between the particles which is translationally invariant.
    \item $\ctrlvec (t)$: The $N$ dimensional control at time $t$. In the present study, $\ctrlvec(t)$ is a continuous function of time, $t$. 
    \item $D$: Diffusion constant.
    \item $\xivec(t)$: $N$ dimensional white noise. The statistics of the white noise is given by, $\langle \mathbf{\xi}_i^{\alpha} \rangle$=0, and $\langle \mathbf{\xi}_i^{\alpha}(t)\mathbf{\xi}_j^{\beta}(t')\rangle=\delta(t-t')\delta^{\alpha \beta}\delta_{i,j}$, where $\alpha=1,2,..d$ and $1\leq (i,j) \leq N$. Here, $d$ is the ambient dimension. In the present study, $d=2$ for all the numerical experiments.
    \item $\ctrlvec_{[0:T]}$: The control $\ctrlvec (t)$ for $0<t<T$, where $T$ is the time-horizon of the control. 
    \item $\mathbb{E}_{\mathbb{Q}_{[0,T]}}$: Expectation over paths, $\mathbb{Q}_{[0,T]}$, given by Eqn. (1) in the main text.
    \item $W$: Total work done in the time horizon $0<t<T$, given by $W=\int_0^T \left\| \ctrlvec \right\|^2(t) dt$.
    \item $\Psi(\xvec(T))$: Terminal penalty.
    \item $\gamma$: The parameter~$\gamma$ captures the weight assigned to the total work done, $W=\int_0^T \left\| \ctrlvec \right\|^2(t) dt$, relative to the terminal penalty, $\Psi(\xvec(T))$.
    
    \item $F(t,\zvec)$: The optimal control value function. Here, $\zvec$ is a point in $\mathbb{R}^{dN}$. In all the numerical experiments, $d=2$.
    \item $\varphi(t, \zvec)$: Related to $F(t,\zvec)$ by the Cole-Hopf transform, $F(t,\zvec) = - \frac{1}{\beta} \log \varphi (t, \zvec)$, where $\beta=\frac{1}{2\gamma \beta}$.
    \item $\mathcal{L}$: Lagrangian needed to convert constrained optimization to unconstrained optimization.
    \item $\lvec(t)$: Adjoint calculated backward in time.
    \item $\sym(M) = \frac{M+M^\top}{2}$
    \item ${\bf x}_i^s$: Location of the $i^{th}$ static particle used to create $V_s$.
    
\end{itemize}

\section{Derivation of the stochastic HJB by Dynamic Programming}
Here we summarize the derivation of the HJB equation following classic texts such as \cite{RB:57,RS:94,RS:86}.  After a time~$t$ has elapsed, the optimal control value function, $F(t,\zvec)$, 
for remaining time interval $[t,T]$ can be written as
\begin{align*}
   &F(t, \zvec) = \min_{\ctrlvec_{[t,T]}} ~\mathbb{E}_{\mathbb{Q}_{\ctrlvec_{[t,T]}}} \left[ \left. \frac{\gamma}{2} \int_t^T \left\| \ctrlvec (t') \right\|^2 dt' + \Psi(\xvec(T)) ~\right|~  \xvec(t) = \zvec \right] \\
    &= \min_{\ctrlvec(t)}  ~\left \lbrace \frac{\gamma}{2} \left\| \ctrlvec(t) \right\|^2 dt \right. \\
        & \quad \left. + \mathbb{E}_{\mathbb{Q}_{\ctrlvec_{t}}} \left[ \min_{\ctrlvec_{[t+dt,T]}}  \mathbb{E}_{\mathbb{Q}_{\ctrlvec_{[t+dt,T]}}} \left[ \left. \frac{\gamma}{2} \int_{t+dt}^T \left\| \ctrlvec(t') \right\|^2 dt' + \Psi(\xvec(T)) \right| \right. \right. \right. \\
        & \qquad \qquad \qquad \qquad \qquad \qquad \qquad \left. \left. \left. \xvec(t+dt) = \zvec -\nabla V(\zvec) dt + \ctrlvec(t) dt + \sqrt{2D} d\Bvec_t \right] \right] \right \rbrace \\
    &= \min_{\ctrlvec(t)} \left \lbrace \frac{\gamma}{2} \left\| \ctrlvec(t) \right\|^2 dt + \mathbb{E}_{\mathbb{Q}_{\ctrlvec_{t}}} \left[ F(t+dt, \zvec -\nabla V(\zvec) dt + \ctrlvec(t) dt + \sqrt{2D} d\Bvec_t ) \right] \right \rbrace
\end{align*}
Note that we have expressed the dynamics equivalently 
in It$\hat{\rm o}$ form as $d\xvec(t) = -\nabla V(\xvec(t)) dt + \ctrlvec(t) dt + \sqrt{2D} d\Bvec_t$ (where $\Bvec_t$ is a standard Brownian motion) above. 
By Taylor expanding to first order in $dt$, we obtain
\begin{align*}
    0 &= \min_{\ctrlvec(t)} \left \lbrace \frac{\gamma}{2} \left\| \ctrlvec(t) \right\|^2 dt 
            +  \frac{\partial F}{\partial t} dt 
                -  \nabla F \cdot \nabla V dt + \nabla F \cdot \ctrlvec(t) dt + D \Delta F dt \right \rbrace \\
    &= \frac{\partial F}{\partial t} + D \Delta F  -  \nabla F \cdot \nabla V
        + \min_{\ctrlvec(t)} \left \lbrace \frac{\gamma}{2} \left\| \ctrlvec(t) \right\|^2  + \nabla F \cdot \ctrlvec(t) \right \rbrace,
\end{align*}
where $\Delta$ is the Laplace operator. The right-hand side above attains a minimum at $\ctrlvec^*(t) = - \frac{1}{\gamma} \nabla F(t, \xvec(t))$, and we get
\begin{align*} 
    \frac{\partial F}{\partial t} + D \Delta F  -  \nabla F \cdot \nabla V - \frac{1}{2\gamma} \left| \nabla F \right|^2 = 0
\end{align*}
along with the boundary condition $F(T,\zvec) = \Psi(\zvec)$.

\section{Derivation of the stochastic HJB equation 
from the Fokker-Planck equation}
Here we summarize the derivation of  the stochastic HJB equation 
from the Fokker-Planck equation following the work of \cite{RJ-DK-FO:98}. Writing the F-P equation as  
\begin{align*}
  \frac{\partial p}{\partial t}(t, \zvec) = D \Delta p(t,\zvec) - \nabla \cdot \left( (- \nabla V(t,\zvec) + \ctrlvec(t,\zvec)) p(t,\zvec) \right) 
\end{align*}
The optimal control cost can be reformulated as
\begin{align*}
    \frac{\gamma}{2} \int_0^T \int_{\Omega} p(t,\zvec) \left\| \ctrlvec(t,\zvec) \right\|^2 ~d\zvec dt + \int_{\Omega} p(T,\zvec) \Psi(\zvec)~d\zvec
\end{align*}
The optimal control problem is then given by
\begin{align*}
    \min_{\ctrlvec}~ &\frac{\gamma}{2} \int_0^T \int_{\Omega} p(t,\zvec) \left\| \ctrlvec(t,\zvec) \right\|^2 ~d\zvec dt + \int_{\Omega} p(T,\zvec) \Psi(\zvec)~d\zvec \\
    &\text{s.t.}~ \frac{\partial p}{\partial t}(t,\zvec) = D \Delta p(t,\zvec) - \nabla \cdot \left( (- \nabla V(t,\zvec) + \ctrl(t,\zvec)) p(t,\zvec) \right) 
\end{align*}
The Lagrangian (let the Lagrange multiplier function be $F(t,\zvec)$ for the constraint) for the above problem is given by
\begin{align*}
    \mathcal{L} = &\frac{\gamma}{2} \int_0^T \int_{\Omega} p(t,\zvec) \left\| \ctrlvec(t,\zvec) \right\|^2 ~d\zvec dt + \int_{\Omega} p(T,\zvec) \Psi(\zvec)~d\zvec \\
    &- \int_0^T \int_{\Omega} F(t,\zvec) \left[ \frac{\partial p}{\partial t}(t,\zvec) - D \Delta p(t,\zvec) + \nabla \cdot \left( (- \nabla V(t,\zvec) + \ctrl(t,\zvec)) p(t,\zvec) \right) \right] ~d\zvec dt
\end{align*}
Optimal control field $\ctrlvec^*(t,\zvec) = - \frac{1}{\gamma} \nabla_{\zvec} F(t,\zvec)$ and the evolution of $F$ is given by (from the first variation with respect to $p$)
\begin{align*}
    - \frac{\partial F}{\partial t}(t,\zvec) 
    = D \Delta F(t,\zvec) - \nabla F (t,\zvec) \cdot \nabla V(t,\zvec)
        - \frac{1}{2\gamma} \left| \nabla F (t, \zvec) \right|^2
\end{align*}
with the terminal condition 
\begin{align*}
    F(T,\zvec) = \Psi(\zvec)
\end{align*}

\section{Derivation of Adjoint-based Path Integral Control (APIC)}
We first recall the path integral~\cite{PDM:04} representation of the optimal value function $F(t, \zvec)$ from the main text
\begin{align*}
    F(t,\zvec) = - \frac{1}{\beta} \log \left( \mathbb{E}_{\mathbb{P}_{[t,T]}} \left[ \left. e^{-\beta \Psi(\xvec(T))} \; \right| \;
        \dot{\xvec}(t') = - \nabla V(\xvec(t')) + \sqrt{2D} \xivec(t'), ~ \xvec(t) = \zvec  \right] \right),
\end{align*} 
where the expectation above is taken with respect to the distribution 
$\mathbb{P}_{[t,T]}$ of paths generated by the \emph{uncontrolled dynamics} $\dot \xvec(t') = -\nabla V(\xvec(t') + \sqrt{2D}\xivec(t')  $ over the time interval $t' \in [t,T]$ starting at $\zvec$, i.e., satisfying the condition $\xvec(t) = \zvec$.
Furthermore, we recall that $F(t,\zvec) = - \frac{1}{\beta} \log \varphi (t, \zvec)$, where $\varphi (t, \zvec)$ is given by
\begin{align*} 
     \varphi(t, \zvec) =  \mathbb{E}_{\mathbb{P}_{[t,T]}} \left[ \left. e^{-\beta \Psi(\xvec(T))}  \; \right| \;
        \dot{\xvec}(t') = - \nabla V(\xvec(t')) + \sqrt{2D} \xivec(t'), ~ \xvec(t) = \zvec \right].
\end{align*}
We see that computing the optimal control 
$\ctrlvec^*(t) = - \frac{1}{\gamma} \nabla F(t, \xvec(t))$ 
at time $t$ involves propagating the state $\xvec(t)$ through the
uncontrolled dynamics over the time interval $[t,T]$ 
to a new state $\xvec(T)$ at which point the gradient is evaluated,
i.e., the gradient is evaluated at a distance.
We accomplish this via the adjoint method, which we formulate below.
From the above path integral representations for $F(t,\zvec)$ and $\varphi(t, \zvec)$, we first note that their gradients are related as 
follows
\begin{align*}
    \nabla F(t,\zvec) 
    = - \frac{\nabla \varphi(t,\zvec)}{\beta \varphi(t,\zvec)} 
\end{align*}
To compute the gradient of $\varphi(t,\zvec)$ with respect to $\zvec$, we note that the uncontrolled dynamics $\dot{\xvec}(t') = - \nabla V(\xvec(t')) + \sqrt{2D} \xivec(t')$ must be propagated starting from $\zvec$ at time $t$ (i.e., $\xvec(t) = \zvec$) over the time interval~$[t,T]$.
We do this by calculus of variations~\cite{RTR:97}, whereby we treat the uncontrolled dynamics as a constraint to construct the Lagrangian for the evaluation of the expectation in~\eqref{eq:FK_formula_varphi}
\begin{align*}
    \mathcal{L}(t, \xvec, \nuvec) =  \mathbb{E}_{\xivec} \left[ e^{-\beta \Psi(\x(T))} - \int_t^T \nuvec^\top \left( \dot{\xvec}(t')  + \nabla V(\xvec(t')) - \sqrt{2D} \xivec(t') \right) dt' \right],
\end{align*}
where $\nuvec$ is the Lagrange multiplier for the constraint of
uncontrolled dynamics and the expectation above is with respect to 
the stochastic process~$\xivec$.
Taking the first variation of the Lagrangian $\mathcal{L}$ 
with respect to $\xvec$, we get
\begin{align*}
    \delta \mathcal{L} &= \mathbb{E}_{\xivec} \left[  -\beta \left. e^{-\beta \Psi(z)} \nabla \Psi(z) \right|_{z = \xvec(T)} \delta \xvec(T) - \nuvec(T) \delta \xvec(T) + \nuvec(t) \delta \xvec(t) \right. \\
    &\qquad \left. + \int_t^T  \dot{\nuvec}(t')^\top \delta \xvec(t')  dt' - \int_t^T \nuvec(t')^\top \nabla^2 V(\xvec(t')) \delta \xvec(t') dt'  \right] \\
    &= \mathbb{E}_{\xivec} \left[ \nuvec(t) \delta \xvec(t) + \left( - \beta  \left. e^{-\beta \Psi(z)} \nabla \Psi(z) \right|_{z = \xvec(T)}  - \nuvec(T) \right) \delta \xvec(T) \right. \\
    &\qquad \left. + \int_t^T \left( \dot{\nuvec}(t') - \nabla^2 V(\xvec(t')) \nuvec(t') \right)^\top \delta \xvec(t') dt' \right]
\end{align*}
From setting the first variation with respect to $\xvec$ in the interval $(0,T)$ and the terminal $\xvec(T)$ above to zero, we obtain the
stationary conditions which fix the evolution of the Lagrange multiplier
$\nuvec$ as follows
\begin{align*}
    &\dot{\nuvec}(t') = \nabla^2 V(\xvec(t')) \nuvec(t'), \\
    &\text{subject~to} \quad \nuvec(T) = - \beta e^{-\beta \Psi(\xvec(T))} \nabla \Psi(\xvec(T)).
\end{align*}
Furthermore, taking the first variation of the Lagrangian~$\mathcal{L}$ with respect to~$\nuvec$, we retrieve the uncontrolled dynamics as the governing equation for~$\xvec$. Once we fix the dynamics of~$\nuvec$ as above (with the uncontrolled dynamics determining $\xvec$), we get $\delta \mathcal{L} = \mathbb{E}_{\xivec} \left[ \nuvec(t) \right] \delta \xvec(t)$.
Since $\xvec(t) = \zvec$, it follows that $\nabla \mathcal{L} (t, \zvec, \nuvec) = \mathbb{E}_{\xivec} \left[ \nuvec(t) \right]$.
Further, we see that the gradient of~$\varphi$ with respect to~$\zvec$ 
is related to the gradient of the Lagrangian~$\mathcal{L}$, and
is given by
\begin{align*}
    \nabla \varphi(t,\zvec) 
    = \nabla \mathcal{L}(t,\zvec,\nuvec) 
    = \mathbb{E}_{\xivec} \left[ \nuvec(t) \right]
\end{align*}
Recalling the relation between the gradients of~$F$ and~$\varphi$, we get
\begin{align*}
    \nabla F(t,\zvec) 
    = - \frac{\nabla \varphi(t,\zvec)}{\beta \varphi(t,\zvec)}
    = - \frac{\mathbb{E}_{\xivec}[\nuvec(t)]}{\eta \mathbb{E}_{\xivec} \left[ e^{-\beta \Psi(\xvec(T))} \right]}
    =  \mathbb{E}_{\xivec} \left[ - \frac{ \nuvec(t) }{\beta \mathbb{E}_{\xivec} \left[ e^{-\beta \Psi(\xvec(T))} \right]  } \right]
\end{align*}
We now make a change of variables 
\begin{align*}
    \lambdavec(t') = - \frac{ \nuvec(t') }{\beta \mathbb{E}_{\xivec} \left[ e^{-\beta \Psi(\xvec(T))} \right]  },
\end{align*}
to obtain the adjoint equations for the gradient computation of the path integral as follows
\begin{align*} 
\begin{aligned}
    \dot{\lambdavec}(t') = \nabla^2 V(\xvec(t')) \lambdavec(t'), \qquad 
    \lambdavec(T) = \frac{e^{-\beta \Psi(\xvec(T))}}{\mathbb{E}_{\xivec} \left[ e^{-\beta \Psi(\xvec(T))} \right]} \nabla \Psi(\xvec(T)), 
\end{aligned}
\end{align*}
where $\nabla^2 V$ is the Hessian of potential $V$ and the optimal control at time~$t$ is then given by
\begin{align*} 
    \ctrlvec^*(t) = - \frac{1}{\gamma} \left. \nabla F(t,\zvec) \right|_{\zvec = \xvec(t)} = - \frac{1}{\gamma} \mathbb{E}_{\xivec} \left[ \lambdavec(t) \right]
\end{align*}

\section{Shooting method for athermal optimal control}
We construct a control Hamiltonian 
$\mathcal{H}(\xvec, \lambdavec) = -\frac{1}{2\gamma} \| \lambdavec \|^2 - \lambdavec^\top \nabla V(\xvec)$ to express the evolution of the optimally controlled system as a Hamiltonian system
\begin{align*}
    \dot{\xvec}(t) &= \frac{\partial \mathcal{H}}{\partial \lambdavec}(\xvec(t), \lambdavec(t)) \\
    \dot{\lambdavec}(t) &= - \frac{\partial \mathcal{H}}{\partial \xvec}(\xvec(t), \lambdavec(t))
\end{align*}
The initial state $\xvec(0) = \xvec_0$ is given
and the initial co-state $\lambdavec(0) = \lambdavec_0$ is chosen to 
minimize the terminal cost $\Psi(\xvec(T))$, via the following shooting problem \cite{MD-SG:11} 
\begin{align*}
    \min_{\lambdavec_0} \Psi(\xvec(T)) \quad \text{s.t.}~ 
    \begin{cases}
        \dot{\xvec}(t) &= \frac{\partial \mathcal{H}}{\partial \lambdavec}(\xvec(t), \lambdavec(t)) \\
    \dot{\lambdavec}(t) &= - \frac{\partial \mathcal{H}}{\partial \xvec}(\xvec(t), \lambdavec(t)) \\
    \xvec(0) &= \xvec_0
    \end{cases}
\end{align*}

\section{Parameters for all the numerical experiments}
\begin{itemize}
    \item \textit{Stochastic optimal control in a frozen landscape of hills and valleys (Figure~3A)}: \\
    The parameters $\gamma = 1$, $D= 5 \times 10^{-4}$ and $T=3$. 
    The static potential was chosen as follows
    \begin{align*}
        V(\mathbf{z}) = \sum_{i=1}^n \eta_i \Theta \left( \left( \left\| \mathbf{z} - \mathbf{r}_i \right\| - d_0 \right)^2 \right)
    \end{align*}
    where $\Theta$ is the Heaviside step function, the number of modes in the frozen landscape $n = 80$,
    $\eta_i$ was sampled uniformly from the interval $(-0.01, 0.01)$, 
    $\mathbf{r}_i$ was sampled uniformly from a square of side $0.8$ units and $d_0 = 0.04$.
    In the numerical simulations, the time step for numerical integration by the Euler-Maruyama method 
    was chosen to be $dt = 0.01$ and the number of sampled paths in the implementation of the Adjoint-based PI control was $20$.
    \item \textit{Stochastic Optimal control of interacting  particles in flat (Figure~5-B,C) and frozen (Figure~3-B,C) landscapes }: \\
    $N=5$ interacting particles, with initial positions sampled uniformly within a square box of size $0.05$ centered at $(-0.4, -0.4)$ and the control task was to steer the particles to the origin $(0,0)$. For the flat landscape case, the parameters were chosen to be $K=0.0001$ and $D=0.01$ in one experiment and $K=1.0$ and $D=0.01$ in another. For the frozen landscape case, a Gaussian mixture landscape was chosen with $20$ modes and covariance $\Sigma = 0.1 I$ ($I$ being the two-dimensional identity matrix). The parameters were chosen to be $D=0.0001$ and $K=0.0001$ in one experiment and $D=0.01$ and $K=0.1$ in another. In the numerical simulations, the time step for numerical integration by the Euler-Maruyama method was chosen to be $dt = 0.01$ and the number of sampled paths in the implementation of the Adjoint-based PI control was $20$.  
    
    \item \textit{Deterministic Optimal control of interacting athermal particles (Figure~4)}: Initial configuration of a non-linear spring network, where the node is a particle's position, and the edge is the non-linear spring interaction. The potential energy, V,  is given by, 
    \begin{align*}
        V({\bf x}_1,...,{\bf x}_N)=\frac{1}{2}\sum_{i=1}^{N}\sum_{j\in NN(i)}k(||{\bf x_i}-{\bf x_j}||-l)^2,
    \end{align*}
 where $N=30$, $k=0.1$, $l=0.2$. The initial $x$ and $y$ coordinates of the particles are sampled from a uniform distribution with support from [-3, 3]. The nearest-neighbor of the $i^{th}$ particle, $NN(i)$, is based on a distance cut-off of 1.5 units, and remains the same throughout the experiment. Three different time horizons, $T$, were chosen as $0.1$, $0.5$, and $1$ time units with the temporal step size $0.00001$ time units. The control task was to go to the circle with radius $4$ units in time, $T$.
    \item \textit{Deterministic optimal control of a single particle in a frozen landscape (Figure~5A)}: Plot of single particles moving on a frozen landscape,
    \begin{align*}
        V_{\rm s}({\bf r})=\frac{\nu}{\sqrt{2\pi \sigma^2}} \sum_{i=1}^{i=N}e^{-\frac{||{\bf x}_i^s-{\bf r}||^2}{2\sigma^2}},
    \end{align*}
     where {\bf r} is a spatial location in two-dimensions, ${\bf x}_i^s$ is the location of the $i^{th}$ static particle, generated from a uniform distribution, which generates the Gaussian potential, $N=20$, $\sigma^2=0.1$, and $\nu=-0.01$. The control task was to move from the initial location (0.1, -0.3), to final point (0, 0). Particle trajectories corresponding to four time-horizons: $T=0.12, T=0.1, T=0.05,$ and $T=0.01$ with temporal step size $0.0000001$ time units.

      \item \textit{Stochastic Optimal control of a particle in a frozen landscape (Figure~S1)}: \\
    $N=5$ non-interacting particles, with initial positions sampled within a square of size $0.05$ units centered at $(-0.4, -0.4)$ and the control task was to steer the particles to the origin $(0,0)$.
    Parameters $D = 10^{-4}$, $\gamma = 10^{-5}$ and $T=5$. In the numerical simulations, the time step for numerical integration by the Euler-Maruyama method was chosen to be $dt = 0.01$ 
    and the number of sampled paths in the implementation of the Adjoint-based PI control was $20$.

\end{itemize}

\section{Details of movies from numerical experiments}
Movies from numerical experiments are accessible at~\cite{SS-VK-LM:23_movies}.

\begin{enumerate}
    \item[\bf Movie1] Stochastic optimal control in a frozen landscape of hills and valleys ({\bf MovFig3A.mp4}). See Figure 3A caption in the main text for details of the movie.
    
    \item[\bf Movie2] Stochastic Optimal control of interacting  particles in frozen landscapes ({\bf MovFig3BC.mp4}). The details for numerics with the frozen landscape can be found in Figure 3B, 3C caption in the main text. 
    
    \item[\bf Movie3] Deterministic Optimal control of interacting athermal particles ({\bf MovFig4.mp4}). Details of the numerics are given in Figure 4 caption in the main text.
    
    \item[\bf Movie4] Deterministic optimal control of a single particle in a frozen landscape ({\bf MovFig5A.mp4}). Details of the numerics are given in Figure 5A caption in the main text.
    
    \item[\bf Movie5] Stochastic Optimal control of interacting  particles in flat landscapes ({\bf MovFig5BC.mp4}). The details for numerics with the frozen landscape can be found in Figure 1B, 1C caption in the main text.  The details for numerics with the flat landscape can be found in Figure 5B, 5C caption in the main text. 

    \item[\bf Movie6] Stochastic Optimal control of a particle in a frozen landscape ({\bf MovFigS1.mp4}). See Figure S1 caption in the SI for details of the movie.
\end{enumerate}

\section{Supplementary figures}
\begin{figure}[H]
    \includegraphics[width=1.0\linewidth]{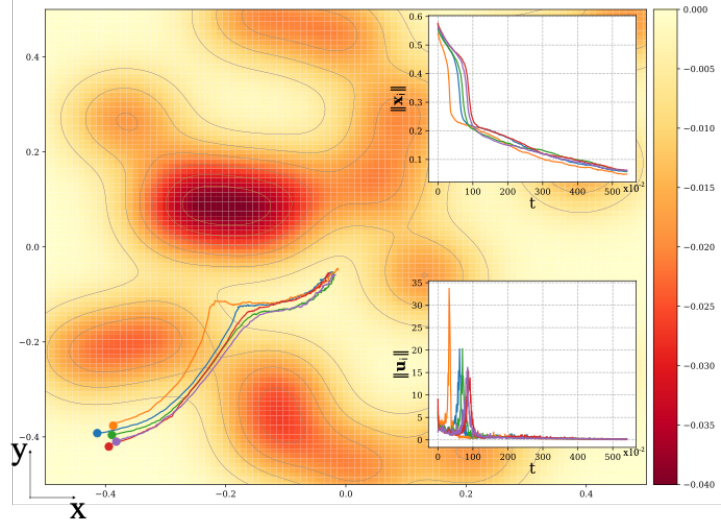}
\caption{{\bf Set of stochastic non-interacting particles in a complex frozen landscape}. The main plot shows the trajectories of five non-interacting particles. The particles were initialized in a square box of size $0.05$ units centered at $(-0.4, -0.4)$ and the control task was to steer the particles to the origin $(0,0)$ while minimizing the total work done. The differently colored dots refer to the initial location of the particles. The frozen complex landscape was implemented using the Gaussian mixture potential (color bars on the right depicts the magnitude of the frozen potential). The trajectories of the particles, shown using curves of different colors, clearly shows the preferential motion along saddles (regions of flat landscape). This behavior is consistent with the theoretical derivation of emergent trajectories given by Eqn. (12) of the main text. The inset in the upper-right corner shows the plot of distance of particles from the origin, $||{\bf x}_i||(t)$. The inset in the lower-right corner shows the plot of the control, $||{\bf u}_i||(t)$. When the particles move along the ridge, there are spikes in the magnitude of control which are correlated with the sudden drop in the distance from origin. }
\label{fig:non_int_stochastic_land}
\end{figure}




\end{document}